\documentclass[prd,showpacs,preprintnumbers,onecolumn,amsmath,nofootinbib,amssymb]{revtex4}
\usepackage{graphicx,color,dcolumn,booktabs,bm}
\usepackage{longtable,lscape}
\usepackage{txfonts}
\usepackage{overpic}
\usepackage{float}
\usepackage{amssymb}
\usepackage{epstopdf}
\usepackage{indentfirst}
\usepackage{feynmf}   %{feynmp}
\usepackage{slashed}  %for Feynman symbolsf
\usepackage{cases}
\usepackage{ulem}
\usepackage{color}
\usepackage{float}
\usepackage{multirow}
\usepackage{graphicx,color,dcolumn,booktabs,bm}
\usepackage{epsfig,dsfont,amssymb,amsmath,amsfonts,amsbsy,mathrsfs}
\usepackage{tikz}
\usepackage{epstopdf}
\usepackage{graphicx,color,dcolumn,booktabs,bm}
\usepackage{epsfig,dsfont,amssymb,amsmath,amsfonts,amsbsy,mathrsfs}

\graphicspath{{Figures/}} %
\usepackage[colorlinks, citecolor=blue,anchorcolor=red,menucolor=red, linkcolor=red,filecolor=red,runcolor=red,urlcolor=blue,frenchlinks=red]{hyperref}
\makeatletter
\@addtoreset{equation}{section}
\makeatother

\newcommand{\nc}{\newcommand}%мӵÄ
\nc{\tj}[1]{\textcolor{red}{Tianjie: #1}}
\allowdisplaybreaks
\begin{document}

% ---------------------------
%\hspace*{5 in} bconfel [28$^{\rm th}$ July 2014] % after proofs
% -----------------------------------------------------------------------------------%
\title{The eigenvalue of  the confined  potential}
%------------------------------------------------------------------------------------
\author{Cheng-Qun Pang$^{1,2}$}%\footnote{Corresponding author}}\email{pcq@qhnu.edu.cn}
\author{Lei Huang$^{3}$}%\email{huanglei80@126.com}
\author{Duo-jie Jia$^{4}$}%\Email{Jiadj@Nwnu.Edu.Cn}
\author{Tian-Jie Zhang$^{5}$}
%\footnote{Corresponding author}}\email{xiangliu@lzu.edu.cn}
 \affiliation{
 $^1$College of Physics and Electronic Information Engineering, Qinghai Normal University, Xining 810000, China\\
  $^2$Joint Research Center for Physics, Lanzhou University and Qinghai Normal University, Xining 810000, China\\
$^3$Heze medical college, Heze 274000, China\\
$^4$ Institute of Theoretical Physics, College of Physics and Electronic Engineering, Northwest Normal University, Lanzhou 730070, China\\
$^5$ School of Mathematics and Statistics, Ningxia University, Yinchuan 750000, China}
\begin{abstract}
\noindent  Analytic solutions for the energy eigenvalues are obtained from a confined potentials of the form $br$ in 3 dimensions. The confinement is effected by linear term which is a very important part in Cornell potential. The analytic eigenvalues and numerical solutions are exactly matched.

\end{abstract}

\pacs{31.15.-p 31.10.+z 36.10.Ee 36.20.Kd 03.65.Ge.}
\maketitle
%%%%%%%%%%%%%%%%%%%%%%%%%%%%%%%%%%%%%%
%\section{Introduction}\label{intro}
%%%%%%%%%%%%%%%%%%%%%%%%%%%%%%%%%%%%%%
%--------------------------------------------------
%--------------------------------------------------
\section{Introduction and the formula }
This paper is concerned with the energy of  system with   the confined  potential (linear)  and obeying nonrelativistic quantum mechanics.
 The main result is  that we obtain an exact formula for the energy eigenvalues.
The confined  potential (linear) is of interest in many quantum mechanics system, such as the Stark effect of atom. Furthermore,
The Cornell potential (Coulomb-plus-linear potential) is very success in describing the spectrum of the quarkonium \cite{McClary:1983xw,Turbiner:1987xa,Bykov:1985it,Ebert:2000bi}.
The eigenvalue of Cornell potential was studied for many years \cite{Hall:1984wk,Papp1993Energy,Guedes2001Solution,Plante2005Analytic,Deloff:2006xx,Casaubon2007Variation,Chung:2008sm,Ovsiyuk:2011mp,Kudryashov:2009sj,Chen:2013hna,Hall:2014xcm,caruso2014solving,Andreev2017}, and yet there is no exact result for the eigenvalues.
References \cite{Hall:1984wk,Ovsiyuk:2011mp,caruso2014solving,Nasser2014Spectra,Andreev2017} gave some approach formulas.
 In this paper, we obtain an exact formula for eigenvalue of the linear part of Cornell potential which can describe the high excited states for the quarkonium \cite{Pang:2018gcn,Pang:2019ttv}.

We consider a Schr\"odinger equation  of the form $(-\frac{1}{2\mu}\Delta +V(r))\psi = E\psi$,
 where $V(r)$ is an attractive central potential with confining terms read:
\begin{equation}
V(r) = b r-\frac{\alpha}{r}-C.
\end{equation}

If we now choose
\begin{equation}
\label{wf}
\psi _{nlm}\left( {\bf r}\right) =\frac{R_{nl}\left(
r\right)}rY_{lm}\left( \theta ,\varphi \right),
\end{equation}
the Schr\"odinger equation will has the form:
\begin{equation}
\label{schr}
\left[ -\frac{d^2}{dr^2}+\frac{l\left( l+1\right)}{r^2}+2\mu\left(b r-\frac{\alpha}{r} -C\right) \right] R_{nl}\left(
r\right)=2\mu E R_{nl}\left( r\right).
\end{equation}

When $\alpha=0$ and $l(l+1)=0$,  it become an Airy equation and has the solution $N_n {\bf{Airy}}((2\mu b)^{1/3}r+ {\bf{AiryZero}}(n+1))$, the eigenvalue will be $\lambda_{n0}=-(2\mu b)^{2/3} {\bf {AiryZero}}(n+1)$, where $N_n$ is the normalized factor and {\bf{AiryZero}}(n) is the {$n$-th} zero point of Airy function.

{Used the transformation of}

\begin{equation}
\label{schrt}
r=\xi/\sigma, a=\frac{2\mu \alpha}{\sigma}, \lambda_{nl}= \frac{2\mu(E+C)}{\sigma^2},  \sigma=(2\mu b)^{1/3},
\end{equation}
the Schr\"odinger equation {in Eq.~\ref{schr}} reads

\begin{equation}
\label{asse}
\left[ -\frac{d^2}{d\xi^2}+\frac{l\left( l+1\right)}{\xi^2}+\left(\xi-\frac{a}{\xi}\right) \right] R_{nl}\left(
\xi\right)=\lambda_{nl} R_{nl}\left(\xi\right).
\end{equation}

We can easily give the series solution of the equation \ref{asse}
\begin{equation}
\label{ri}
R_{nl}(\xi)=\sum_i c_{i}\xi^{i+l+1}
\end{equation}
and the coefficient $c_i$ reads:

\begin{equation}
\label{ci}
c_i=(-1)^{i} \frac{\bf Det A }{\bf Det B}c_0,(i>0){,}
\end{equation}

\begin{equation}
\label{Aii}
A_{pq}=\delta_{p+1}^{q}p(p + 1 + 2 l) +\delta_{p}^{q}a+\delta_{p-1}^{q}\lambda_{nl}- \delta_{p-2}^{q},\\
B_{pq}=\delta_{p}^{q}p(p + 1 + 2 l) +\delta_{p-1}^{q}a+\delta_{p-2}^{q}\lambda_{nl} - \delta_{p-3}^{q}.
\end{equation}
And we can obtain that

%\begin{eqnarray}
%$
% \KK_{i=0}^m \left( \frac{a_i}{b_i} \right) =
%  \cfrac{a_0}{
%  b_0 + \cfrac{a_1}{
%  b_1 + \genfrac{}{}{0pt}{0}{}{\displaystyle\ddots + \cfrac{a_m}{b_m}}}}, \qquad
% \KK_{i=0}^\infty \left( \frac{a_i}{b_i} \right) =
%  \cfrac{a_0}{
%  b_0 + \cfrac{a_1}{
%  b_1 + \genfrac{}{}{0pt}{0}{}{\ddots }}}.$
%\end{eqnarray}

%\begin{eqnarray}
%\label{bii}
% \KK_{i=0}^m \left( \frac{a_i}{b_i} \right) =
%  \cfrac{a_0}{
%  b_0 + \cfrac{a_1}{
%  b_1 + \genfrac{}{}{0pt}{0}{}{\displaystyle\ddots + \cfrac{a_m}{b_m}}}}, \qquad
% \KK_{i=0}^\infty \left( \frac{a_i}{b_i} \right) =
%  \cfrac{a_0}{
%  b_0 + \cfrac{a_1}{
%  b_1 + \genfrac{}{}{0pt}{0}{}{\ddots }}}.
%\end{eqnarray}

\begin{equation}
{\bf Det A}={\bf{NF}}[a+\underset{s=2}{\overset{i}{\mathrm{\mathbf{K}}}}\frac{a_s}{a}],
\end{equation}
where NF representivs take numerator and

\begin{equation}
a_i=- \lambda (i-1) (i+2 l),
\end{equation}\begin{equation}
 a_{i-1}=-(i-2) (2 l+i-1) \left(\frac{ (i-1) (2 l+i)}{a}+\lambda_{nl} \right),
\end{equation}
\begin{equation}
a_{i-k}=-(i-k-1) (i-k+2 l) \left(\frac{ (i-k) (i-k+2 l+1)}{\underset{s=i-k+2}{\overset{i}{\mathrm{\mathbf{K}}}}\frac{a_s}{a}+a}+\lambda_{nl}\right),
\end{equation}

\begin{equation}
\underset{i=1}{\overset{n}{\mathrm{\mathbf{K}}}}\frac{a_i}{a}=\frac{a_1}{a+\frac{ a_2}{a+\frac{a_3}{a}\text{...}}},
\end{equation}
and

\begin{equation}
\label{bii}
{\bf Det B}=i!\Pi_{j=1}^{i}(2l+j+1),
\end{equation}
{where $Det(A)$ and $Det(B)$  are denoted by the determinants of matrixes $A$ and $B$ respectively}.
When we select a eigenvalue of Eq. \ref{asse}, the series \ref{ri} is convergent.

%------------------------------------------------------------
%\section{a formula for the eigenvalue of  confined  potential}
%------------------------------------------------------------
We give an ansatz that the  eigenvalue of Eq. \ref{asse}, has the following form with $a=0$,
\begin{equation}\label{lambda}
\lambda_{nl}=-\frac{\delta _1 l+\frac{\pi ^{2/3}  }{\sqrt[3]{3}}\delta _2n+1 }{\frac{\pi ^{2/3}}{\sqrt[3]{3}}(\delta _1 l+ \delta _2n)+1}{\bf{AiryZero}}(l+n+1).
\end{equation}
And we find that when $\delta_1=0.797533$, $\delta_2=0.797804$, $\delta_1\approx\delta_2\approx {\sqrt{\frac{2}{\pi}}}\approx 0.798\approx 0.8\approx\delta$, which is solved by the numerical eigenvalues of $\lambda_{11} $, $\lambda_{01}=3.361254522$,  $\lambda_{02}=4.88445124014$ with $a=0$ and $b=1$. This eigenvalues formula  and numerical solutions are {exactly} matched as illustrated in Tabs. \ref{table:tab1} and  \ref{table:tab2}. For the $\lambda_{0l}$, the analytic eigenvalues and numerical solutions are exact {exactly} matched, and for the radial excitation,  the relative error between analytic eigenvalues and numerical results below 0.0004. As shown in Tabs. \ref{table:tab1} and  \ref{table:tab2}, the results of WKB approach is only well described the low $l$ excitation and our formula is consistent with the numerical eigenvalues of linear potential for all $n$ and $l$. We also calculate the $\lambda_{100~100}$ and $\lambda_{0~100}$ numerically and by using the formula \ref{lambda}, and obtain  41.08626 and 41.08631, 80.6248  and 80.650, respectively.

\begin{table}[!h] \caption{The comparison between this work and few numerical results by using simple harmonic oscillator (SHO) basis  expansion(we use 20 SHO bases) or other method. } %title of Table
\[\begin{array}{c c c c}
\hline
\hline
\begin{array}{ccccc}
\multicolumn{3}{c}{\lambda_{0l},a=0} \\
  l& \text{This work} &\text{WKB}&\text{ Numerical} \\
  \hline
 0 & 2.33811&2.32025 &2.33811 \\
 1 & 3.36125 &3.26163& 3.36125 \\
 2 & 4.24819 &4.08181& 4.24818 \\
 3 & 5.05095 &4.82632& 5.05093 \\
 4 & 5.79446 & 5.51716&5.79442 \\
 5 &6.49306 &6.16713& 6.49303\\
 6 & 7.15596 &6.78445& 7.15589 \\
 7 & 7.78947 &7.37485& 7.78942 \\
 8 &8.39822 & 7.94249&8.39817 \\
 9 & 8.98566 &8.49051& 8.98561\\
 10 &9.55449&9.02137& 9.55443\\
% 11 &10.1069 & 10.1068\\
% 12 &10.6446&10.6445 \\
% 13 & 11.169&11.1689 \\
% 14 & 11.6814& 11.6813\\
\end{array}
 &
\begin{array}{ccc}
\multicolumn{3}{c}{\lambda_{0l}, a=1} \\
  l& \text{This work} &\text{ Numerical } \\
    \hline
  0 & 1.39750& 1.397876 $\cite{Hall:2014xcm}$ \\
 1 & 2.82564 & 2.82565  $\cite{Hall:2014xcm}$ \\
 2 & 3.85138& 3.85058  $\cite{Hall:2014xcm}$ \\
 3 & 4.72713 & 4.72675 $\cite{Hall:2014xcm}$ \\
 4 & 5.51702 & 5.51698 $\cite{Hall:2014xcm}$ \\
 5 & 6.24822& 6.24840 $\cite{Hall:2014xcm}$ \\
 6 &6.93556& 6.93583\\
 7 & 7.58827 & 7.58852 \\
 8 & 8.21255& 8.21274\\
 9 & 8.81287& 8.81297 \\
 10 &9.39258& 9.39260\\
\end{array}
 &
\begin{array}{ccc}
\multicolumn{3}{c}{\lambda_{n0},a=1} \\
  n& \text{This work} & \text{ Numerical} \\
    \hline
  0 & 1.39749 & 1.397876 $ \cite{Hall:2014xcm}$  \\
 1 & 3.48278 & 3.475087  $\cite{Hall:2014xcm}$ \\
 2 & 5.14824 & 5.032914  $\cite{Hall:2014xcm}$ \\
 3 & 6.39233 & 6.370149  $\cite{Hall:2014xcm}$ \\
 4 & 7.60665 &7.574933 $\cite{Hall:2014xcm}$ \\
 5 & 8.72888& 8.687915  $\cite{Hall:2014xcm}$ \\
 6 & 9.78119& 9.7360 \\
 7 & 10.7780& 10.7261 \\
 8 & 11.7292 & 11.6749 \\
 9 & 12.6421 & 12.5835 \\
 10 & 13.5221&13.4712 \\
\end{array}
 \\
 \hline
\hline
\end{array}\]
\label{table:tab1}
\end{table}

\begin{table}[!h] \caption{The comparison between this work and few numerical result by using simple harmonic oscillator (SHO) basis  expansion(we use 20 SHO bases) for  $a=0$. The numerical results is consist with the Ref. \cite{Chen:2013hna,Deloff:2006xx,Hall:2014xcm}}. %title of Table
\[\begin{array}{ccc}
\hline
\hline
\begin{array}{cccccc}
  n& l&\text{This work}&\text{WKB} &\text{ Numerical} \\
  \hline
 1 & 1 & 4.88445 &4.82632& 4.88445 \\
 2 & 1 & 6.20822 &6.16713& 6.20762 \\
 3 & 1 & 7.40683 &7.37485& 7.40567 \\
 4 & 1 & 8.51685 &8.48051& 8.51523 \\
 5 & 1 & 9.55959 & 9.53705&9.55762 \\
 6 & 1 & 10.5488 &10.5290& 10.5465 \\
 7 & 1 & 11.4939 & 11.4762&11.4914 \\
 8 & 1 & 12.4019 & 12.3857&12.3992 \\
 9 & 1 & 13.2779 & 13.2630&13.2751 \\
 10 & 1 & 14.126 &14.1122& 14.1232 \\
 11 & 1 & 14.9495 & 14.9366&14.9467 \\
 12 & 1 & 15.751 &15.7388& 15.7481 \\
 13 & 1 & 16.5326&16.5210 & 16.5309 \\
 14 & 1 & 17.2962 & 17.2852&17.2948 \\
\end{array}
 &
\begin{array}{ccccc}
  n& l&\text{This work}&\text{WKB}  &\text{ Numerical} \\
  \hline
 1 & 8 & 9.41274&9.02137 & 9.41302 \\
 2 & 8 & 10.3821 & 10.0391&10.3822 \\
 3 & 8 & 11.3128 &11.0077 &11.3125 \\
 4 & 8 & 12.2103 &11.9353& 12.2094 \\
 5 & 8 & 13.0785 &12.8281& 13.0769 \\
 6 & 8 & 13.9207 & 13.6909&13.9185 \\
 7 & 8 & 14.7398 &14.5273 &14.7369 \\
 8 & 8 & 15.5381 &15.3403 &15.5344 \\
 9 & 8 & 16.3173 &16.1323 &16.313 \\
 10 & 8 & 17.0791 &16.9053 &17.0742 \\
 11 & 8 & 17.8251 & 17.661&17.8196 \\
 12 & 8 & 18.5563 & 18.4008&18.5502 \\
 13 & 8 & 19.2738 & 19.1261&19.2681 \\
 14 & 8 & 19.9787 &19.8379 &19.9728 \\
\end{array}
 &
\begin{array}{cccccc}
  n& l&\text{This work}&\text{WKB} &\text{ Numerical} \\
  \hline
  1 & 14 & 12.5481 &11.9353&12.5484 \\
 2 & 14 & 13.3901 &12.8281  & 13.3905 \\
 3 & 14 & 14.2098 &13.6909  &14.21 \\
 4 & 14 & 15.0093 &  14.5273&15.0092 \\
 5 & 14 & 15.7903 &15.3403 & 15.7897 \\
 6 & 14 & 16.5544 &16.1323  &16.5533 \\
 7 & 14 & 17.3027 &16.9053  &17.3011 \\
 8 & 14 & 18.0366 &  17.661&18.0343 \\
 9 & 14 & 18.7569 &18.4008  &18.754 \\
 10 & 14 & 19.4646 & 19.1261 &19.4612 \\
 11 & 14 & 20.1606 &19.8379  &20.1565 \\
 12 & 14 & 20.8455 &20.5371  &20.8408 \\
 13 & 14 & 21.5199 & 21.2246 &21.5153 \\
 14 & 14 & 22.1845 &21.9012 & 22.1796 \\
\end{array}
 \\
 \hline\hline
\end{array}\]
\label{table:tab2}
\end{table}
\section{Application }
\subsection{Regge trajectory and the WKB approach of linear potential }
{Using} this eigenvalue formula in Eq. \ref{ri}, one can obtain the eigenfunctions. In the other hand, this is {the} important step to reach the exact solution to Eq. \ref{asse}.
%\section{The Aplication of the eigenvalue formula}
We can expand the Eq. \ref{lambda} when $n+l+1\rightarrow \infty$ with

\begin{equation}\label{nla}
\lambda_{nl}=\frac{(12 \pi )^{2/3}}{4} \frac{\left(l+n+\frac{3}{4}\right)^{2/3} \left(\delta  \left(l+\frac{1}{3} (3 \pi )^{2/3} n\right)+1\right)}{\frac{1}{3} (3 \pi )^{2/3} \delta  (l+n)+1}
\end{equation}

In fact, Eq. \ref{nla} has the high precision for all $n$ and $l$.

%\begin{equation}
%\label{nl1}{\frac{3}{2^{2/3}} \left(\frac{\pi ^{2/3}}{\sqrt[3]{3}}-1\right) {{(l+n)}^{-1/3}}}n+3 \left(\frac{l+n}{2}\right)^{2/3}
%\end{equation}

We can expand the Eq. \ref{lambda} when $n\gg l$ with
\begin{equation}
\label{n}
\lambda _{{n0}}^3\approx\left(\frac{3 \pi}{2}n\right)^2,
\end{equation}
or
\begin{equation}
\label{nl}
\lambda _{{n0}}^3\approx\left(\frac{3 \pi}{2}(n+l)\right)^2,
\end{equation}
which indicate that when $n$ or $l$ is bigger enough, $n+l$
can be regard as the principle quantum number consist with the guess of Ref. \cite{Glozman:2007ek,Glozman:2007at}, and the similar formula was obtained in Ref.
\cite{Chen:2019uzm} from the relativistic flux tube model.
When $l\gg n$, the approximate  will be
\begin{equation}
\label{l}
\lambda _{{0l}}^3\approx\left(\frac{3^{3/2} }{2}l\right)^2.
\end{equation}
or
\begin{equation}
\label{ln}
\lambda _{{0l}}^3\approx\left(\frac{3^{3/2} }{2}(l+n)\right)^2.
\end{equation}
Eqs \ref{n} and \ref{l} are the famous results for the linear potential \cite{Chen:2018hnx}.

\subsection{The eigenvalue of  the conell  potential }
In this section, {we will give }an explicit form  for  the Schr\"odinger equation \ref{asse}.
The  eigenvalue of  the Coulomb   potential reads

\begin{equation}
\label{bii}
\lambda(a) _{{nl}}=-\frac{a^2}{4(n+l+1)^2}.
\end{equation}

So, we ansatz the eigenvalue of  the conell  potential has the form:

\begin{equation}
\label{abf}
\lambda _{{nl}}=-\frac{a^2}{4(n+l+1)^2}-\frac{\delta _1 l+\frac{\pi ^{2/3} }{\sqrt[3]{3}} \delta _2n+1 }{\frac{\pi ^{2/3}}{\sqrt[3]{3}}(\delta _1 l+ \delta _2n)+1}{\bf{AiryZero}}(l+n+1)f(a,n,l)+g(a),
\end{equation}
where $f(0,n,l)=1,g(0)=0$.
Then we try the following form:
\begin{equation}
\label{fx}
f(a,n,l)=1-\frac{ \left(\frac{{al_1}+{al_2} ({an_1} n+l)^{3/2}+{al_3} ({an_2} n+l)^3}{{al_4} ({an_1} n+l)^{3/2}+{al_5} ({an_2} n+l)^3+1}+\frac{{bl_1}+{bl_2} \sqrt{{bn_1} n+l}+{b_3} ({bn_2} n+l)}{{bl_4} \sqrt{{bn_1} n+l}+{bl_5} ({bn_2} n+l)+1}a^{2/5} \right)a^{2/5} }{ \left(\frac{{cl_1}+{cl_2} ({cn_1} n+l)^2+{cl_3} ({cn_2} n+l)^4}{{cl_4} ({cn_1} n+l)^2+{cl_5} ({cn_2} n+l)^4+1}+\frac{ {dl_1}+{dl_2} ({dn_1} n+l)+{dl_3} ({dn_2} n+l)^2}{{dl_4} ({dn_1} n+l)+{dl_5} ({dn_2} n+l)^2+1}a^{2/5}\right)a^{2/5}+\frac{{en_1} n+1}{{en_2}n+1}},
g(a)=f_1(1 - e^{-f_2a})
\end{equation}
with \

\begin{equation}
\begin{aligned}
\label{g}
{al}_1=-0.0131096,{al}_2=-0.0526298,{al}_3=0.000656495,{al}_4=5.04779,{al}_5=1.59813,\\
{bl}_1=0.214579,{bl}_2=-0.0439848,
{bl}_3=0.00362465,{bl}_4=-0.22573,{bl}_5=0.839639,\\
{cl}_1=-0.689677,
{cl}_2=-1.91553,{cl}_3=-0.274089,{cl}_4=4.19812,{cl}_5=0.75866,\\
{dl}_1=0.365051,
{dl}_2=0.148248,{dl}_3=0.0362142,{dl}_4=1.43633,{al}_5=0.621341,\\
{an}_1=0.254949,{an}_2=0.500001,{bn}_1=1.29783,{bn}_2=0.77,\\
{cn}_1=0.0400669,{cn}_2=1.61998,{dn}_1=0.433161,{dn}_2=1.92501,\\
{en}_1=0.4,{en}_2=0.520996,{f}_1=0.00685891,{f}_2=8.47589.
\end{aligned}
\end{equation}
By using {Eq.~}\ref{abf}, we can obtain the eigenvalues of Cornell potential with the relative error is below 0.03 in the ranges of $0\leq a\leq 5$ and $n,l\leq 10$, below 0.001 for n=0.

\renewcommand{\arraystretch}{1.2}
\begin{table}[htbp]
\caption{Bottomnium($ \Upsilon(nS_1)$) mass spectrum using formula \ref{abf}. The parameters of Cornell potential are $b$=0.18 GeV$^2$, $C$=0.29 GeV, $\alpha=0.52$, and $\mu= \frac{4.93}{2}$ GeV. The mass has the unit of GeV. \label{bb}}
\begin{center}
\begin{tabular}{ccccc}
\toprule[1pt]\toprule[1pt]
\text{State} &  This work &Ref. \cite{Faustov:1997ue}&\text{Numerical}&\text{Experiment \cite{Beringer:1900zz}} \\
 \midrule[1pt]
$0^3S_1$&9.4225&9.447&9.4222&9.4603\\
$1^3S_1$&10.013&10.012&10.055&10.023\\
$2^3S_1$&10.360&10.353&10.350&10.355\\
$3^3S_1$&10.641&10.629&10.628&10.579\\
$4^3S_1$&10.889&--&10.871&10.882\\
$5^3S_1$&11.114&--&11.092&11.003\\
%$1^3P_1$&9.9096&9.900&9.90910&9.899\\
%2P&10.270&10.260&10.264&10.260\\
%3P&10.552&10.544&10.547&10.512\\
%$1^3D_2$&10.158&10.155&10.158&10.164\\
%$2^3D_2$&10.457&10.448&10.451&--\\
\bottomrule[1pt]\bottomrule[1pt]
\end{tabular}
\end{center}
\end{table}
In Tab. \ref{bb}, we use  \ref{abf} to calculate the spectrum of Bottomnium($ \Upsilon(nS_1)$), and find that it is consistent with Ref. \cite{Hall:2014xcm}, numerical result and the experimental value \cite{Beringer:1900zz} in the case of $a$=2.67.

\subsection{Conclusion}In this work, we  {obtain} an analytic formula  for the energy eigenvalues from a confined potentials of the form $br$ in 3 dimensions which is widely used to solve the spectrum of the heavy meson family. The analytic eigenvalues and numerical solutions are exact matched which is very close to  the exact energy eigenvalues. This can encourage us to find the   exact solution for the Schr\"odinger equation with confined potentials.
\par
The analytic formula  for the energy eigenvalues of linear potential can reproduce the Regge trajectory of its higher n-excitation and l-excitation
\par
By using the analytic formula  for the energy eigenvalues of linear potential, we give an fit for the Cornell potential which is very well to describe its energy eigenvalues. Tested by the Bottomnium($ \Upsilon(nS_1)$) mass spectrum, we find that it is very useful to analyse the spectrum problem.

\section{ACKNOWLEDGMENTS}
This work is supported  by the National Natural Science Foundation of China under Grants No. 11965016, No. 11861051 and No. 11565023, the projects funded by Science and Technology Department of Qinghai Province No. 2018-ZJ-971Q.
\bibliographystyle{apsrev4-1}
\bibliography{hep}

\end{document}